\documentclass[
  journal=pasa,
  manuscript=research-paper, 
  year=2024,
  volume=41,
]{cup-journal}

\usepackage[colorlinks=true,citecolor=blue]{hyperref}
\usepackage{booktabs}

\title{Hierarchical hub-filament structures and gas inflows on galaxy-cloud scales}

\author{J. W. Zhou}
\affiliation{Max-Planck-Institut f\"{u}r Radioastronomie, Auf dem H\"{u}gel 69, 53121 Bonn, Germany}
\email{jwzhou@mpifr-bonn.mpg.de}

\author{Timothy A. Davis}
\affiliation{Cardiff Hub for Astrophysics Research and Technology, School of Physics and Astronomy, Cardiff University, Queens Buildings, Cardiff CF24 3AA, UK}


\begin{document}

\begin{abstract}
We investigated the kinematics and dynamics of gas structures on galaxy-cloud scales in two spiral galaxies NGC5236 (M83) and NGC4321 (M100) using CO (2$-$1) line.
We utilized the FILFINDER algorithm on integrated intensity maps for the identification of filaments in two galaxies. Clear fluctuations in velocity and density were observed along these filaments, enabling the fitting of velocity gradients around intensity peaks. The variations in velocity gradient across different scales suggest a gradual and consistent increase in velocity gradient from large to small scales, indicative of gravitational collapse,
something also revealed by the correlation between velocity dispersion and column density of gas structures. 
Gas structures at different scales in the galaxy may be organized into hierarchical systems through gravitational coupling.
All the features of gas kinematics on galaxy-cloud scale are very similar to that on cloud-clump and clump-core scales studied in previous works. Thus, the interstellar medium from galaxy to dense core scales presents
multi-scale/hierarchical hub-filament structures. 
Like dense core as the hub in clump, clump as the hub in molecular cloud, now
we verify that cloud or cloud complex can be the hub in spiral galaxies. 
Although the scaling relations and the measured velocity gradients support the gravitational collapse of gas structures on galaxy-cloud scales, the collapse is much slower than a pure free-fall gravitational collapse.
\end{abstract}


\section{Introduction} 

Measuring the dynamical coupling between density enhancements in giant molecular clouds and gas motion of their local environment gives us a perspective to understand the formation of hierarchical structures in high-mass star formation regions \citep{McKee2007-45, Motte2018-56, Henshaw2020-4}. 
High-resolution observations of high-mass star-forming regions reveal the structured arrangement of density enhancements within filamentary gas networks, notably in hub-filament systems. Within these systems, converging flows channel material into the central hub through the interconnected filaments
\citep{Peretto2013,Henshaw2014,Zhang2015,Liu2016,Yuan2018,Lu2018,Issac2019,Dewangan2020,Liu2022-511,Zhou2022-514,Zhou2023-676}. In observations, gas inflow will naturally produce velocity gradients along filaments \citep{Kirk2013,Liu2016,Yuan2018,Williams2018-613,Chen2019-875,Chen2020-891,Pillai2020-4,Zhou2022-514,Zhou2023-676}.

\citet{Henshaw2020-4} identified widespread velocity fluctuations spanning various spatial scales and physical contexts within galaxies. They observed oscillatory gas movements with wavelengths ranging from 0.3 to 400 parsecs, intricately linked to regularly spaced density enhancements. These enhancements are likely a result of gravitational instabilities
\citep{Henshaw2016-463, Elmegreen2018-863}. Furthermore, the spatial correlation between density enhancements and velocity gradient extrema may indicate convergent motion due to gravitational collapse \citep{Hacar2011-533, Hacar2016-587, Clarke2016-458, Misugi2019-881, Zhou2022-514, Zhou2023-676}. 

\citet{Zhou2022-514} investigated the physical properties and evolution of hub-filament systems in $\sim$ 140 protoclusters, utilizing spectral lines observed in the ATOMS (ALMA Three-millimeter Observations of Massive Star-forming regions) survey \citep{Liu2020}. Their findings indicate the presence of hub-filament structures across a range of scales, spanning from 0.1 parsec to several parsecs, in diverse Galactic environments. Additionally, slender structures resembling filaments, such as spiral arms, have been identified at scales below 0.1 parsecs in the vicinity of high-mass protostars \citep{Liu2015-804,Maud2017-467,Izquierdo2018-478,Chen2020-4,Sanhueza2021-915}. As proposed by \citet{Zhou2022-514}, self-similar hub-filament structures and filamentary accretion seem to exist across scales, ranging from several thousand astronomical units to several parsecs, within high-mass star-forming regions.
This paradigm of hierarchical/multi-scale hub-filament structures was generalized from clump-core scale to cloud-clump scale in \citet{Zhou2023-676}. 
Hierarchical collapse and hub-filaments structures feeding the central regions are also  described in previous works, see \citet{Motte2018-56,Vazquez2019-490, Kumar2020-642} and references therein.

Kinematically, the results in \citet{Zhou2023-676} also reveal the multi-scale hub-filament structures in the G333 molecular cloud complex. 
The G333 complex exhibits prevalent kinematic characteristics consistent with hub-filament systems. Notably, the intensity peaks, acting as hubs, correlate with converging velocities, indicating that the surrounding gas flows are converging to dense structures. Specifically, there is a discernible increase in velocity gradients at smaller scales.
The filaments in the Large APEX sub-Millimeter Array (LAsMA) and the Atacama Large Millimeter/submillimeter Array (ALMA) observations show clear velocity gradients. The velocity gradients fitted using the LAsMA and ALMA data exhibit consistent agreement over the range of scales covered by ALMA observations in the ATOMS survey ($\textless$ 5\,pc). In \citet{Zhou2023-676}, larger scale gas motions were investigated (the longest filament $\sim$50\,pc), yet similar results were obtained compared to small-scale ALMA observations. 
Interestingly, the variations in velocity gradients measured at distinct scales—small scales ($\textless$ 1\,pc), medium scales ($\sim$ 1-7.5\,pc), and large scales ($\textgreater$ 7.5\,pc)—align with expectations from gravitational free-fall with central masses of $\sim$ 500\,M$_\odot$, $\sim$ 5000\,M$_\odot$ and $\sim$\, 50000\,M$_\odot$ for the respective scales. 
This implies that sustaining velocity gradients on larger scales necessitates higher masses. 
The association of higher masses with larger scales suggests that the inflow on a larger scale is driven by the larger-scale structure which may be the gravitational clustering of smaller-scale structures. This observation aligns with the hierarchical nature commonly found in molecular clouds and the gas inflow from large to small scales. 
The large-scale gas inflow is likely driven by gravity, indicating a state of global gravitational collapse within the molecular clouds in the G333 complex. This supports the argument that these molecular clouds serve as cloud-scale hub-filament structures.
The change in velocity gradients with the scale in the G333 complex indicates that the morphology of the velocity field in position-position-velocity (PPV) space resembles a "funnel" structure. 
The funnel structure can be elucidated as the acceleration of material converging towards the central hub and the gravitational contraction of star-forming clouds or clumps. 
Large-scale velocity gradients always associate with many intensity peaks, and the larger-scale inflow is driven by the larger-scale structure, indicating that the clustering of smaller-scale gravitational structures locally can serve as the gravitational center on a larger scale. In a way, the funnel structure provides insight into the gravitational potential well shaped by this clustering.

In previous work, we have investigated gas dynamics and kinematics in molecular clouds from cloud to core scales. The main task of this work is to generalize the above physical pictures to galaxy-cloud scales, now the smallest unit is the molecular cloud itself. In \citet{Zhou2022-514}, dense core is the hub in clump, in \citet{Zhou2023-676}, clump is the hub in molecular cloud, here we treat molecular cloud as the hub in galaxy. Self-similar or hierarchical/multi-scale hub-filament structures and filamentary accretion feeding the central regions exist from molecular cloud to dense core scales \citep{Motte2018-56,Vazquez2019-490, Kumar2020-642, Zhou2022-514,Zhou2023-676}, this picture will be extended to galaxy-cloud scales in this work.

\section{Data and target}
We select two face-on spiral galaxies NGC5236 (M83) and NGC4321 (M100) from the PHANGS-ALMA survey. We use the combined 12m+7m+TP PHANGS-ALMA CO (2$-$1) data
cubes to investigate gas kinematics and dynamics in the two galaxies, which have a spectral resolution of 2.5 km s$^{-1}$ and angular resolutions $\sim$2.1 $''$ and $\sim$1.7 $''$, corresponding
to linear resolutions $\sim$51 pc and $\sim$123 pc for NGC5236 and NGC4321 at the distances 4.89 Mpc and 15.21 Mpc \citep{Leroy2021-257,Anand2021-501}, respectively. The field-of-view (FOV) of NGC5236 and NGC4321 in the ALMA observations are $\sim$ 10.5 Kpc $\times$ 10.8 Kpc and $\sim$ 17.1 Kpc $\times$ 15.5 Kpc, respectively.
An overview of the PHANGS-ALMA survey's science goals, sample selection, observation strategy, and data products is described in \citet{Leroy2021-257}. A detailed description of data calibration, imaging, and product creation procedures is presented in \citet{Leroy2021-255}. The
PHANGS-ALMA CO (2$-$1) data cubes and other data products (such as Moment maps) are available from
the PHANGS team website \footnote{\url{https://sites.google.com/view/phangs/home}}. NGC5236 and NGC4321 were selected for the following reasons:

1. They are face-on galaxies. The inclination angles of NGC5236 and NGC4321 are 24$^{\rm o}$ and 38.5$^{\rm o}$ \citep{Lang2020-897}, respectively. Thus, we can ignore the projection effect in velocity gradient fitting in Sec.\ref{gradient0};

2. They have strong CO (2$-$1) emission, presenting large-scale continuous structures (strong spiral arms), thus we can trace galaxy-scale gas motions;

3. NGC5236 has similar morphology to NGC4321, but it is about 3 times closer than NGC4321, thus we can resolve more detailed structures. 

\section{Results}
\subsection{Velocity component}\label{component}
\begin{figure*}
\centering
\includegraphics[width=1\textwidth]{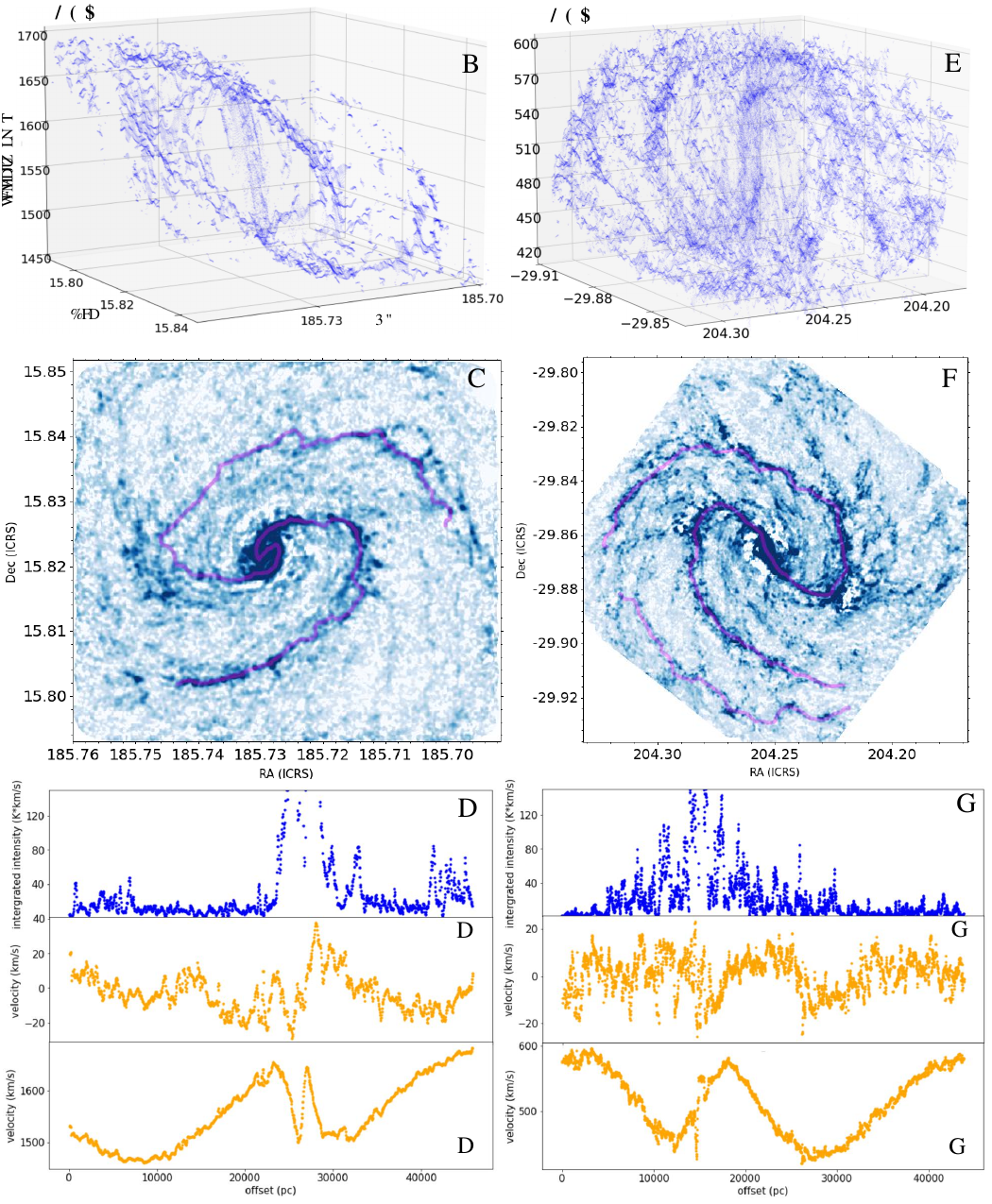}
\caption{ An overview of gas kinematics in NGC4321 and NGC5236.
(a) and (d) The velocity distribution of the entire galaxies in PPV space decomposed by \texttt{GAUSSPY+}; 
(b) and (e) The main filaments of NGC4321 and NGC5236 identified by FILFINDER algorithm overlaid on the Moment 0 maps; 
(c) and (f) The integrated intensity distributions extracted from the Moment 0 maps along the main filaments in panels (b) and (e); 
(c2) and (f2) The velocity distributions extracted from the Moment 1 maps along the main filaments in panels (b) and (e);
(c1) and (f1) The velocity residual distributions along the main filaments in panels (b) and (e) after subtracting the large-scale velocity gradients due to the galaxy rotation.}
\label{3Dv}
\end{figure*}

Before studying gas kinematics, we need to ascertain the distribution of gas components in the galaxy. 
We utilized the fully automated Gaussian decomposer \texttt{GAUSSPY+} \citep{Lindner2015-149, Riener2019-628}, designed to break down intricate spectra of molecular lines into multiple Gaussian components. The parameter settings for the decomposition align with those employed by \citet{Zhou2023-676}.
For the CO (2$-$1) data cube of NGC4321, 50071 spectra were fitted by \texttt{GAUSSPY+}, 98\% spectra are single component, 2\% spectra have two velocity components. 323059 spectra were fitted for NGC5236, 87\% spectra show single component, 11.5\% spectra have two components. The reduced $\chi^{2}$ values output in the fitting of NGC4321 and NGC5236 are around 1, which means that both of them get a good fit. Generally, multiple velocity components ($>$1 velocity component) mainly concentrate on the central region of the galaxy. In this work, we mainly focus on the gas kinematics on the spiral arms. Therefore, the analysis can be carried out based on the Moment maps.

In Fig.\ref{3Dv}, we can see ubiquitous velocity fluctuations, which may be attributed to gravitational instability, as discussed below. In the central region of the galaxy, there is almost no velocity fluctuations, clear velocity fluctuations are mainly distributed on the spiral arms, revealing the interconnection between velocity fluctuations and star formation activities. 

\subsection{Scaling relations}
\subsubsection{Dendrogram} \label{ds}

\begin{figure}
\centering
\includegraphics[width=0.96\textwidth]{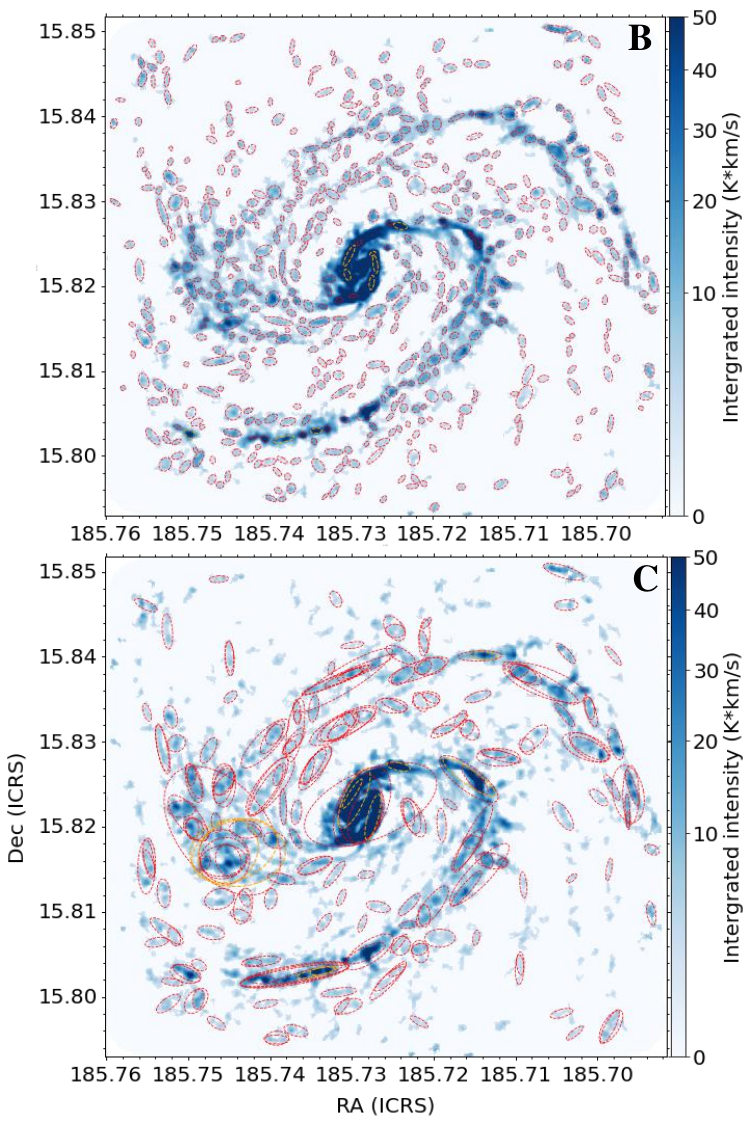}
\caption{Dendrogram structures. (a) Leaves; (b) Branches. Red and orange eclipses represent type1 and type3 structures, respectively.}
\label{dendro1}
\end{figure}

\begin{figure}
\centering
\includegraphics[width=0.96\textwidth]{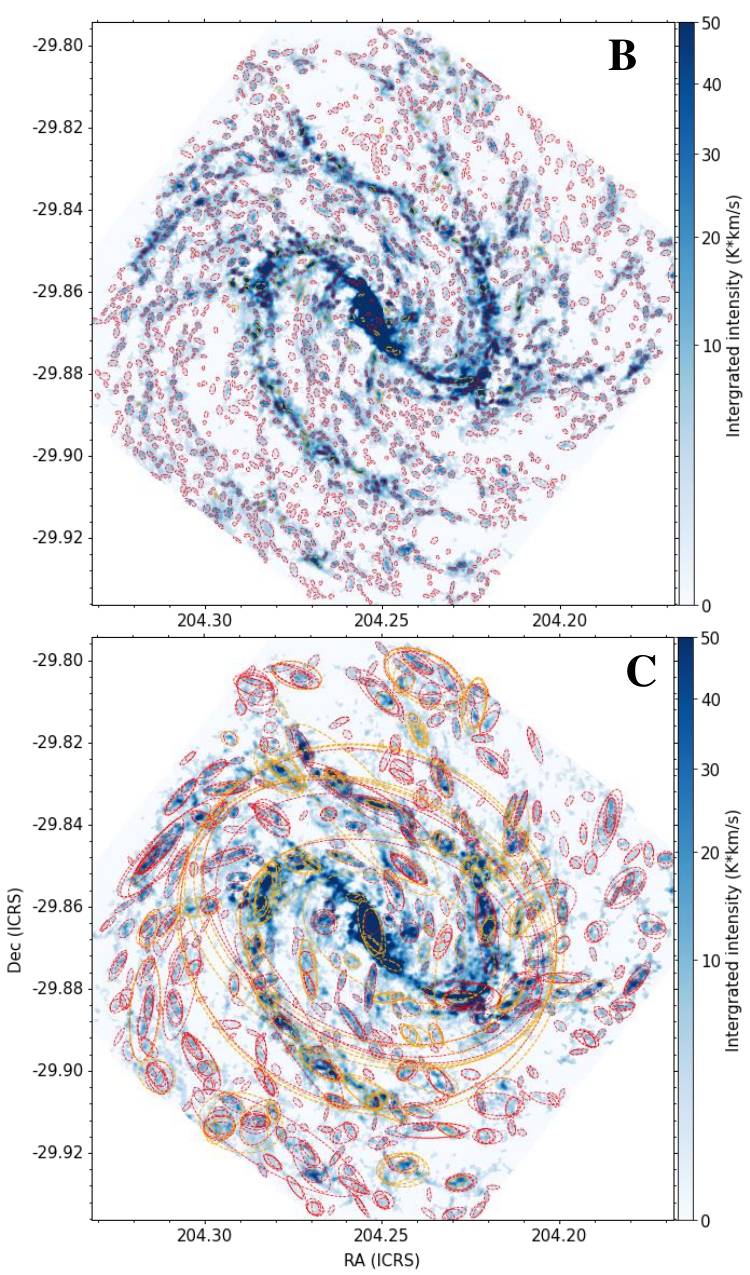}
\caption{Same as Fig.\ref{dendro1}, but for NGC5236.}
\label{dendro2}
\end{figure}

\begin{table}
	\centering
	\caption{Dendrogram structures in NGC 4321 and NGC 5236.}
	\label{line}
	\begin{tabular}{cccccc}
		\hline
	&	leaves	&	branches	&	type1	&	type3 & total	\\
NGC 4321	&	583	&	270	&	828	&	25 & 853	\\
NGC 5236	&	1722	&	859	&	2339	&	242 & 2581\\
       \hline
	\end{tabular}
    \label{structure}
\end{table}

The issues of identifying structures in a PPV cube using the dendrogram algorithm are discussed in \citet{Zhou2024-682-128}. Consequently, we employed the same approach as outlined in \citet{Zhou2024-682-128} and \citet{Zhou2024-682-173} to identify dense gas structures. We conducted a direct identification of hierarchical (sub-)structures based on the 2D integrated intensity (Moment 0) map of CO (2$-$1) emission. Subsequently, we extracted the average spectrum of each identified structure to delve into its velocity components and gas kinematics.
All the retained structures on the strictly masked Moment 0 map are reliable, we therefore only require the smallest area of the identified structure larger than 1 beam and do not set other parameters in the algorithm to reduce the dependence of the identification on the parameter settings. Dendrogram algorithm decomposes the intensity data (Moment 0 map) into hierarchical structures. 
As illustrated in Fig.1 of \citet{Zhou2024-682-128}, all identified structures can be divided into three categories, i.e. i-leaves (isolated leaves), c-leaves (clustered leaves) and branches. In this work, we merged c-leaves and i-leaves. Finally, there are two kinds of structures, i.e. leaf and branch structures. 
In Fig.\ref{dendro1} and Fig.\ref{dendro2}, 
the structures identified by the dendrogram algorithm exhibit a strong correspondence with the background integrated intensity maps.

The algorithm characterizes the morphology of each structure by approximating it as an ellipse. Within the dendrogram, the rms sizes (second moments) of the intensity distribution along the two spatial dimensions define the long and short axes of the ellipse, denoted as $a$ and $b$. As described in \citet{Zhou2024-682-128}, a smaller ellipse is obtained with $a$ and $b$, necessitating a multiplication by a factor of two to appropriately enlarge the ellipse.
Then the effective physical radius of an ellipse is $R\rm_{eff}$ =$\sqrt{2a \times 2b}*D$, where $D$ is the distance of the galaxy.
For a structure with the area $A$, the total integrated intensity of the structure is $I_{\rm CO}$, then
the mass of the structure can be calculated by
\begin{equation}
M = \alpha^{2-1}_{\rm CO} \times I_{\rm CO} \times A,
\label{mass}
\end{equation}
here $\alpha^{2-1}_{\rm CO} \approx 6.7 \rm M_{\odot} \rm pc^{-2} (\rm K*km s^{-1})^{-1}$ \citep{Leroy2022-927}. 

The large-scale velocity gradients due to the galaxy rotation will contribute to the non-thermal velocity dispersion.
Before extracting the average spectra of the identified structures, we subtracted the large-scale velocity gradients in PPV data cube based on the constructed gas dynamical model in Sec.\ref{kinms}. According to their averaged spectra, the structures with absorption features were eliminated firstly. Following \citet{Zhou2024-682-173}, we only consider type1 (single velocity component) and type3 (blended velocity components) structures in this work.

As detailed in \citet{Zhou2024-682-128}, two screening criteria were applied for structure refinement:
(1) Removal of repetitive branch structures; and
(2) Exclusion of branch structures exhibiting complex morphology.
Finally, there are
853 and 2581 retained structures for NGC4321 and NGC5236, respectively, as listed in Table \ref{structure} and marked in Fig.\ref{dendro1} and Fig.\ref{dendro2} by ellipses. NGC5236 is $\sim$3 times nearer than NGC4321, the total number of the identified structures in NGC5236 is also $\sim$3 times more than that in NGC4321.

\subsubsection{Scaling relations}\label{scaling}
\begin{figure*}
\centering
\includegraphics[width=1\textwidth]{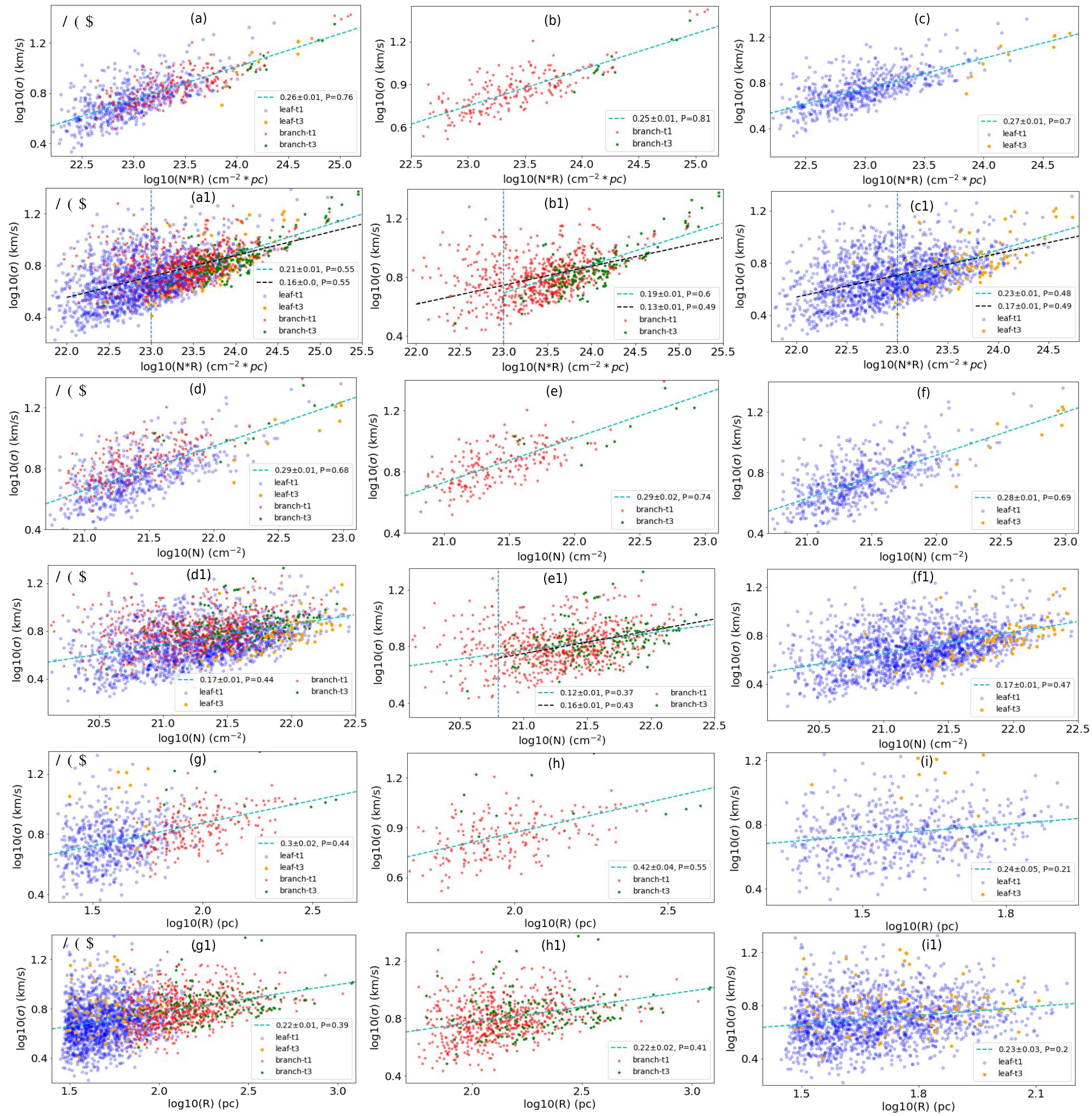}
\caption{Scaling relations of leaf and branch structures of NGC 4321 and NGC 5236. (a)-(c) and (a1)-(c1) $\sigma-N*R$;
(d)-(f) and (d1)-(f1) $\sigma-N$;  
(g)-(i) and (g1)-(i1) $\sigma-R$. $\sigma$, $R$ and $N$ are the velocity dispersion, effective radius, and column density of each structure, respectively. 
Here, ``P'' represents the Pearson coefficient.
Dashed vertical line marks the boundary of two fittings.}
\label{scale}
\end{figure*}

The scaling relations provide insights into the physical states of the structures.
Although the identified structures in NGC5236 are much more than that in NGC4321, the scaling relations in Fig.\ref{scale} are comparable. 
Due to the further distance, some faint structures were filtered out in NGC4321, which can be seen in NGC5236. In Fig.\ref{scale}, the faint or low-density structures produce dispersive tails, similar to the Fig.14 of \citet{Zhou2024-682-128} for the G333 molecular cloud complex in the Milky Way. 

In Fig.\ref{scale}(g)-(i), it appears that only branch structures demonstrate a discernible correlation between velocity dispersion and scale. Clumps or cores characterized by high column density exhibit greater velocity dispersion relative to those with lower column density, a phenomenon attributed to gas motions associated with gravitational collapse
\citep{Ballesteros2011-411,Traficante2018-473,Ballesteros2018-479,Li2023-949,Zhou2024-682-128}.  
As depicted in Fig.\ref{scale}(d)-(f), the positive correlations observed between velocity dispersion and column density suggest a gravitational origin for the velocity dispersion. In the context of pure free-fall, an anticipated $\sigma-N*R$ relation would exhibit a slope of 0.5. For a more convenient comparison with $\sigma-R$ and $\sigma-N$ relations, we transform the Heyer relation, $\sigma/R^{0.5} \propto N^{0.5}$, to the form $\sigma \propto (R*N)^{0.5}$ \citep[Eq.\,3 in][]{Ballesteros2011-411}. Both relations should ideally have a slope of 0.5. However, in Fig.\ref{scale}(a)-(c), the slopes of the $\sigma-N*R$ relations are notably shallower than 0.5, suggesting a deceleration from the expected behavior of pure free-fall gravitational collapse.

\subsection{Velocity gradient} \label{gradient0}

The gravitational collapse of the structures has been revealed by the scaling relations. This section gives a more detailed discussion of the gravitational collapse from the perspective of the velocity gradient.

\subsubsection{Identification of filaments}
The FILFINDER algorithm \citep{Koch2015} was used to identify and characterize filaments in the galaxy. Following the method described in \citet{Zhou2022-514} and \citet{Zhou2023-676}, we used the Moment 0 maps of CO (2$-$1) emission from the data products of the PHANGS-ALMA survey to identify filaments. 
Fig.\ref{3Dv}(b) and (e) display the skeletons of the identified filaments superimposed on the Moment 0 maps. Notably, these skeletons exhibit a high degree of agreement with the gas structures traced by CO (2$-$1) emission.
In this work, we only consider the filaments along the spiral arms (the main filaments), where have strong CO (2$-$1) emission, which allows us to study large-scale continuous gas motions in the galaxy.

Almost all algorithms have the issue of parameter settings, thus we try not to discuss the identified filament itself in this work, such as its length and width.
Actually, we mainly regard the FILFINDER algorithm as a tool to draw lines, these lines can make connections between discrete dense structures, or put dense structures in gas networks. Then by extracting the velocity information along these lines, we can study how the surrounding gas converges to dense gas structures (gravitational centers). 
In \citet{Zhou2022-514} and \citet{Zhou2023-676}, the filaments traverse multiple local hub-filament structures, characterized by fluctuations in velocity and density along the filaments. These local dense structures, acting as gravitational centers, facilitate the accretion of surrounding diffuse gas, culminating in the formation of local hub-filament structures. In the PPV space, a hub-filament structure can be represented as a funnel structure, as demonstrated in Fig.9 of \citet{Zhou2023-676}. The gradient of the funnel profile serves as an indicator of the gravitational field's strength. Consequently, examining the local velocity gradients along the filaments provides insights into the strength of local gravitational fields.

In Fig.\ref{3Dv}, although the velocity fluctuations at small scales can be seen everywhere, they are hidden in global large-scale velocity gradients due to the galaxy rotation.
To derive the local velocity fluctuations, we must first subtract the large-scale velocity gradients.
We take two methods to solve this issue.

\subsubsection{Linear fitting} \label{linear}
\begin{figure}
\centering
\includegraphics[width=0.96\textwidth]{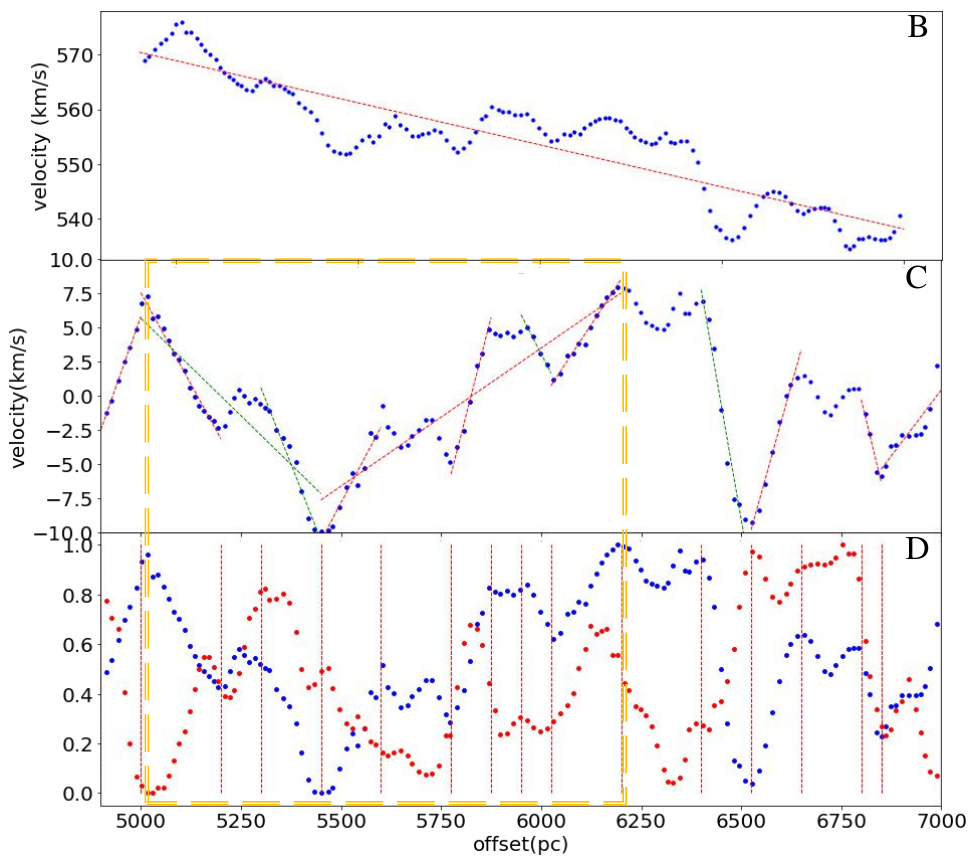}
\caption{A segment of NGC5236's main filament, used to demonstrate the velocity gradient fitting. (a) Linear fitting of the large-scale velocity gradient due to the bulk motion; (b) Velocity field of the segment in panel (a) after subtracting the fitted large-scale velocity gradient. Local velocity gradients are fitted in the ranges defined by the red vertical dashed lines in panel (c), and straight lines show the linear fitting results. The color-coding of straight lines are meaningless, but it is convenient for the reader; (c) Blue and red dotted lines show the normalized velocity and intensity, respectively. Orange box marks
a gravitational coupling structure.
}
\label{fit}
\end{figure}

Fig.\ref{3Dv}(c2)\&(c) and (f2)\&(f)
show the intensity-weighted velocity (Moment 1) and integrated intensity (Moment 0) of CO (2$-$1) line emission along the main filaments with intense velocity and density fluctuations.
A segment of NGC5236's main filament was used to demonstrate the velocity gradient fitting in Fig.\ref{fit}. 
Here, we eliminated the large-scale velocity gradient by employing a simple linear function for modeling. Subsequently, we subtracted this model from the velocity field obtained from the Moment 1 map along the main filament.
We divided the main filaments into many segments to ensure that each segment is as straight as possible, thus increasing the accuracy of the linear fit, as shown in Fig.\ref{fit}(a).
The fitted straight line can be the baseline, then we subtracted the baseline for each segment to leave only the residual local velocity fluctuations. 
The good correspondence between velocity gradients and intensity peaks in Fig.\ref{fit}(b) indicates that the surrounding gases are converging to the center of the gravitational potential well.  
For each segment of the main filament, we firstly estimated the global velocity gradients between velocity peaks and valleys at two sides of the intensity peaks
(representing gravitational centers), and ignored local velocity fluctuations. Then we also derived additional velocity gradients over smaller distances around the local intensity peaks, as illustrated in Fig.\ref{fit}(b). Generally, large-scale velocity gradients are associated with multiple intensity peaks.
Given that NGC4321 and NGC5236 are good face-on galaxies, here we ignored the projection effect in velocity gradient fitting. 

\subsubsection{Gas dynamical model} \label{kinms}
\begin{figure}
\centering
\includegraphics[width=0.96\textwidth]{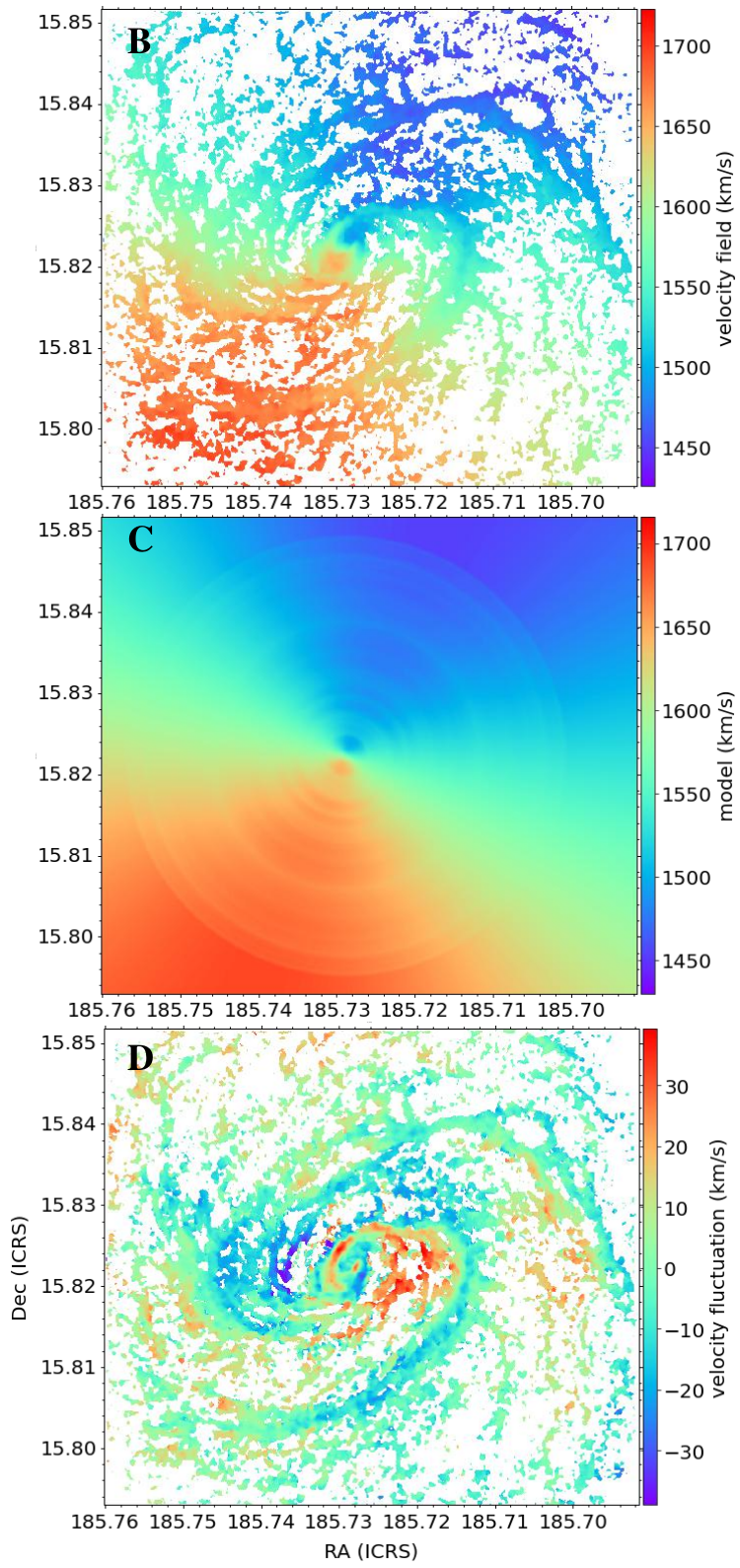}
\caption{(a) The velocity field (Moment 1) of NGC 4321 in observation; (b) Gas dynamical model created by the Kinematic Molecular Simulation (KinMS) package; (c) Velocity fluctuations after removing the created model.}
\label{model}
\end{figure}

As a comparison, we also removed the bulk motion or the galaxy rotation by creating gas dynamical models using the Kinematic Molecular Simulation (KinMS) package of \citet{Davis2013-429} based on the Moment-1 map, as shown in Fig.\ref{model}(b). Then we used the Moment 1 map to subtract the created model, and obtained the velocity residual map. Finally, we extracted the velocity fluctuations from the velocity residual map rather than the Moment 1 map. The subsequent velocity gradient fitting is the same with Sec.\ref{linear}.

\subsubsection{Statistical analysis of the fitted velocity gradients} 

\begin{figure}
\centering
\includegraphics[width=0.96\textwidth]{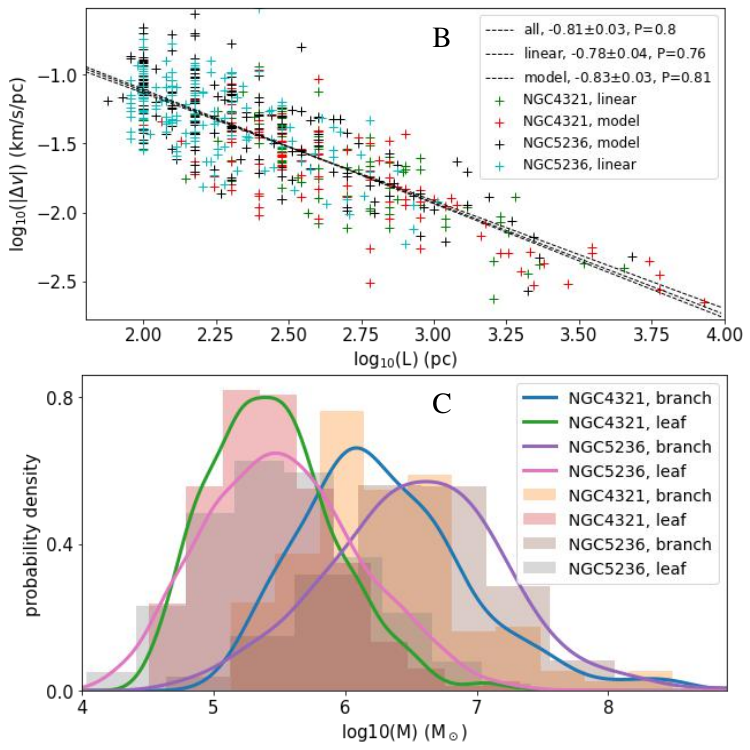}
\caption{(a) The correlation between the fitted velocity gradient and the scale; (b) The mass distribution of the identified structures.}
\label{compare}
\end{figure}
\begin{figure}
\centering
\includegraphics[width=0.96\textwidth]{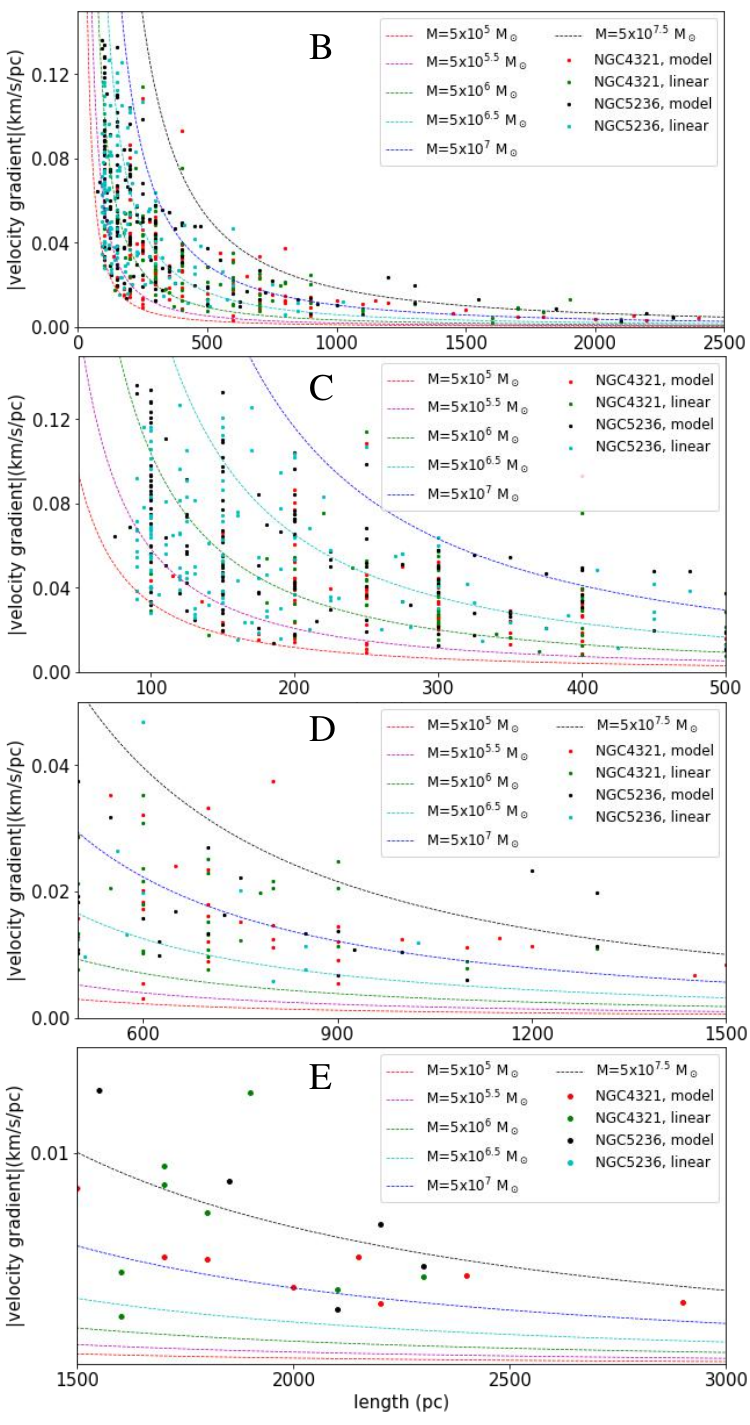}
\caption{Statistical analysis of all fitted velocity gradients.
(a) Velocity gradient vs. the length. The color lines show the freefall velocity gradients for comparison. For the freefall model, red, magenta, green, cyan, blue and black lines denote masses of 10$^{5}$\,M$_\odot$, 10$^{5.5}$\,M$_\odot$, 10$^{6}$\,M$_\odot$ , 10$^{6.5}$\,M$_\odot$, 10$^{7}$\,M$_\odot$ and 10$^{7.5}$\,M$_\odot$, respectively. Panels
(b), (c), and (d): Zoomed maps with lengths $\textless$ 500\,pc (small scale), $\sim$ 500 -- 1500\,pc (medium scale), and $\textgreater$ 1500\,pc (large scale) in panel (a).}
\label{gradient}
\end{figure}

In Sec.\ref{linear} and Sec.\ref{kinms},
the galaxy centers were excluded in the fitting. As shown in Fig.\ref{compare}, two methods give consistent fitting results. 
In Fig.\ref{gradient}, we can find the same gas kinematic modes presented in \citet{Zhou2023-676} and \citet{Zhou2024-682-173}.
The variations in velocity gradients at both small and large scales (with a boundary around 500 pc) align with expectations from gravitational free-fall, with central masses of $\sim$ 10$^{5}$--10$^{6.5}$ \,M$_\odot$ and $\sim$ 10$^{6}$--10$^{7.5}$ \,M$_\odot$. This implies that sustaining velocity gradients on larger scales requires correspondingly larger masses, and larger masses imply larger scales, suggesting that the larger-scale inflow is driven by the larger-scale structure which may arise from the gravitational clustering of smaller-scale structures, in harmony with the presence of hierarchical or multi-scale hub-filament structures within the galaxy and the gas inflows from large to small scales.
In the orange box marked in Fig.\ref{fit}(b) and (c), multiple peaks are coupled together to form a gravitational potential well on larger scale, and each peak itself is also a local gravitational potential well. 

In Fig.\ref{compare}(a), 
almost all measured velocity gradients can be fitted in the mass range $\sim$ 10$^{5}$--10$^{7}$ \,M$_\odot$, which is consistent with the mass distribution of 
the identified structures shown in Fig.\ref{compare}(b), indicating that local dense structures and their complex as gravitational centers will accrete the surrounding diffuse gas and then produce the observed velocity gradients at different scales.

\subsubsection{Deviation of the free-fall model}\label{free}
We can do a deeper analysis for the fitted velocity gradients with a simple model.
Assuming free-fall, 
\begin{equation}
\nabla v = -\frac{d}{dR}\sqrt{\frac{2GM}{R}}
=\sqrt{\frac{GM}{2R^3}},
\label{e-free}
\end{equation}
in this equation, velocity gradient is more sensitive to scale. Thus we only fit the correlation between velocity gradient and scale in Fig.\ref{compare}(a). In equation.\ref{e-free}, $\nabla v \propto R^{-1.5}$, however, the linear fitting in Fig.\ref{compare}(a) gives $\nabla v \propto R^{-0.8}$, thus $\sim$2 times smaller than the slope of the free-fall model,
also indicating the slowing down of a pure free-fall gravitational collapse presented in Sec.\ref{scaling}.

\section{Discussion}

Gas kinematics on galaxy-cloud scales clearly deviate from the free-fall model. The deviation may come from measurement or calculation biases. We only measured the projection of the realistic velocity vector and hence acceleration, thus tends to underestimate the observed acceleration. 
Moreover, we only considered the molecular line CO (2$-$1), which only traces part of the baryonic matters in the galaxy, thus  underestimated the mass of the gravitational center.  

The shallower slope in Fig.\ref{compare}(a) may have physical correlation with the flat rotation curve of the galaxy. Gas motions on galaxy-cloud scales may still couple with the galactic potential, 
a comprehensive model should account for the interplay between motions within the galactic potential and the self-gravitational potential of the cloud.
\citep{Dobbs2013-432,Meidt2018-854,Meidt2020-892,Utreras2020-892}. 
In the model proposed by \citet{Meidt2020-892}, the transition to free-fall collapse can only happen once the gas has completely decoupled from the galactic potential.
However, it is not clear down to which spatial scales gas motions remain dynamically relevant to galactic potential or start decoupling with galactic potential. 

Tidal forces have been advocated in previous works to restrict or trigger star formation. 
Previous studies by \citet{Ballesteros2009-393,Ballesteros2009-395} have shown the impact of tidal forces arising from an effective Galactic potential on molecular clouds. These forces have the potential to either compress or disrupt molecular clouds, influencing the overall star formation efficiency. \citet{Thilliez2014-31} examined the stability of molecular clouds within the Large Magellanic Cloud (LMC) against shear and the galactic tide. However, their findings indicate that star formation in the LMC is not impeded by either tidal or shear instability.
Moreover, in \citet{Ramirez2022-515}, tidal stresses from neighbouring molecular cloud complexes may increase interstellar turbulence, rather than the galactic potential. 

Gravity is a long-range force.
A local dense structure evolves under its self-gravity, but as a gravitational center, 
its influence can also affect neighboring structures. At the same time, it also experiences the gravitational pull from other nearby sources.
The tidal and gravitational fields are mutually interdependent. 
As described above, the hierarchical/multi-scale gravitational coupling of gas structures means the extensive tidal interactions in the galaxy.
Whether the galactic potential has effect on molecular clouds and their complexes or not, molecular clouds should be affected by the cumulative tidal interactions of many nearby materials, which may prevent gravitational collapse and growth of instabilities or star formation in the cloud.
Due to the diffuse and complex morphology of matter distribution in the galaxy, a complete tidal calculation would be very complex. One should derive the gravitational potential distribution from the observed density distribution and then calculate the tidal field according to the gravitational potential distribution, as presented in \citet{Li2024-528}. 
Diverse manifestations of gravitational effects on gas within molecular clouds were unveiled in \citet{Li2024-528}: Dense regions experience gravitational collapse, while the surrounding gas is subject to significant tidal forces, suppressing fragmentation. This gas, influenced by extensive tides, is directed towards the dense regions, serving as fuel for star formation. The spatial distribution of regions experiencing varying tidal influences elucidates the observed hierarchical and localized pattern of star formation.
Similar mechanisms may also exist on galaxy-cloud scales, we will discuss this topic in detail in future work.

\begin{figure}
\centering
\includegraphics[width=0.96\textwidth]{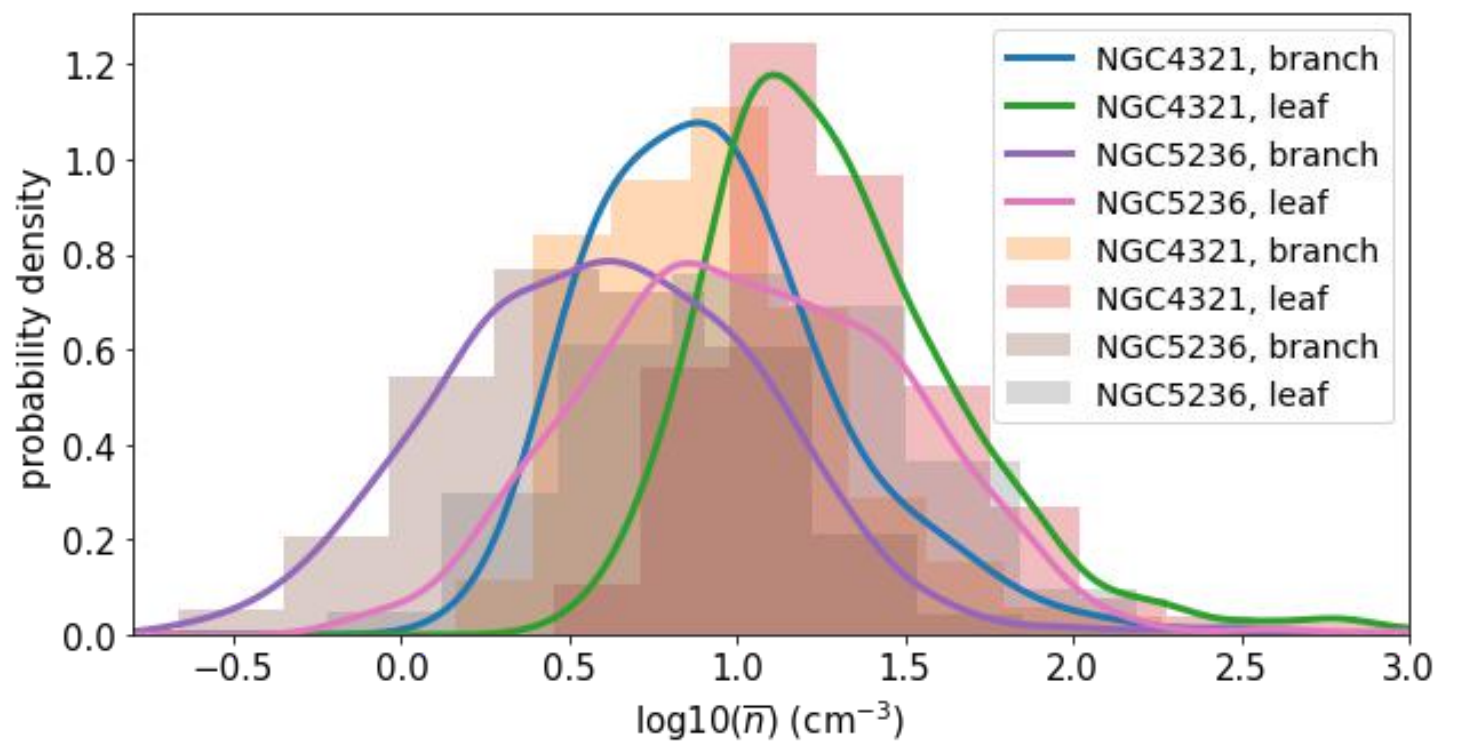}
\caption{The average number density distribution of the identified structures.}
\label{number}
\end{figure}

In addition to the factors discussed above, magnetic fields may have a significant impact on the gas kinematics of molecular clouds as suggested by simulations \citep{Kim2021-911,Seifried2020-497, Ibanez2022-925, Ganguly2023-525} and observations \citep{Crutcher2010-725,Li2011-479,Crutcher2012-50,Stephens2022-926,Ngoc2023-953,Rawat2024-528}. 
Especially, in simulations, 
the diffuse gas with a number density ($n$) less than 100 cm$^{-3}$ in the envelopes of molecular clouds may be upheld against gravitational collapse due to magnetic support \citep{Ibanez2022-925, Ganguly2023-525}. Assuming the spherical geometry of the identified clouds, the average number density can be calculated by
\begin{equation}
\overline{n} = M/(\frac{4}{3} \pi R_{\rm eff}^3)/(\mu m_{\rm H}),
\label{mass}
\end{equation}
where $\mu$ = 2.37 is the mean molecular weight per ‘free particle’ (H$_{2}$ and He, the number of metal particles is negligible), m$_{\rm H}$ is atomic hydrogen mass. As shown in Fig.\ref{number}, almost all of structures have the average number density < 100 cm$^{-3}$. However, the real cloud shape should be more sheetlike instead of spherical \citep{Shetty2006-647,Inutsuka2015-580,Arzoumanian2018-70,Kohno2021-73,Arzoumanian2022-660,Rezaei2022-930,Zhou2023-519,Zhou2023-676,Clarke2023-519,Ganguly2023-525}. Therefore, the average number density of the clouds should be significantly underestimated. Even if the average number density estimates differ by an order of magnitude, what revealed by Fig.\ref{number} remains promising.
The supportive role of the magnetic field may be a significant factor contributing to the deviation from free fall motions, and further detailed investigation is warranted in future research.

\section{Summary}

We investigated the kinematics and dynamics of gas structures on galaxy-cloud scales in two spiral galaxies NGC5236 (M83) and NGC4321 (M100) using the CO (2$-$1) line.
The main conclusions are as follows:

1. We directly identified hierarchical (sub-)structures according to the 2D integrated intensity (Moment 0) map of CO (2$-$1) emission. Subsequently, we extracted the average spectrum for each structure, delving into its velocity components and gas kinematics.
Considering that
the large-scale velocity gradients due to the galaxy rotation will contribute to the non-thermal velocity dispersion, before extracting the average spectra of the identified structures, we subtracted the large-scale velocity gradients in PPV data cube based on the constructed gas dynamical model. 

2. In examining the scaling relation among velocity dispersion ($\sigma$), effective radius ($R$), and column density ($N$) across all structures, it becomes evident that $\sigma-N*R$ consistently exhibits a stronger correlation compared to $\sigma-N$ and $\sigma-R$. The observed correlations between velocity dispersion and column density suggest a potential link to gravitational collapse, corroborated by the measured velocity gradients. However, it is noteworthy that the slopes of the $\sigma-N*R$ relations appear to be significantly shallower than the anticipated value of 0.5, implying a deceleration from the characteristic behavior of a pure free-fall gravitational collapse.

3. We employed the FILFINDER algorithm to identify and characterize filaments within the galaxy using integrated intensity maps. Observable velocity and density fluctuations along these filaments enabled us to fit local velocity gradients around intensity peaks, a process performed after removing the global large-scale velocity gradients attributed to the galaxy's rotation.

4. Statistical analysis of the fitted velocity gradients on galaxy-cloud scales shows the same gas kinematic modes presented on cloud-clump and clump-core scales. 
The variations in velocity gradients at both small and large scales (with a boundary around 500 pc) align with expectations from gravitational free-fall, with central masses of $\sim$ 10$^{5}$--10$^{6.5}$ \,M$_\odot$ and $\sim$ 10$^{6}$--10$^{7.5}$ \,M$_\odot$. This implies that sustaining velocity gradients on larger scales requires correspondingly larger masses, and larger masses imply larger scales, suggesting that the larger-scale inflow is driven by the larger-scale structure which may arise from the gravitational clustering of smaller-scale structures, in harmony with the presence of hierarchical or multi-scale hub-filament structures within the galaxy and the gas inflows from large to small scales.

5. In free-fall model, the velocity gradient and scale satisfy $\nabla v \propto R^{-1.5}$. However, in the observation, $\nabla v \propto R^{-0.8}$, also indicating the slowing down of a pure free-fall gravitational collapse.

\begin{acknowledgement}
J. W. Zhou thanks V. Kalinova for the comments.
It is a pleasure to thank the PHANGS team, the data cubes and other data products shared by the team make this work can be carried out easily. This paper makes use of the following ALMA data: 

ADS/JAO.ALMA\#2013.1.01161.S, 

ADS/JAO.ALMA\#2015.1.00121.S, 

ADS/JAO.ALMA\#2015.1.00956.S, 

ADS/JAO.ALMA\#2016.1.00386.S, 

ADS/JAO.ALMA\#2017.1.00886.L.

ALMA is a partnership of ESO (representing its member states), NSF (USA) and NINS (Japan), together with NRC (Canada), NSTC and ASIAA (Taiwan), and KASI (Republic of Korea), in cooperation with the Republic of Chile. The Joint ALMA Observatory is operated by ESO, AUI/NRAO and NAOJ.

\end{acknowledgement}

\section*{Data Availability}
The PHANGS-ALMA CO (2$-$1) data cubes and other data products (such as moment maps) are available from the PHANGS team website \footnote{\url{https://sites.google.com/view/phangs/home}}.

\bibliography{main}

\end{document}